\documentclass[aip,graphicx]{revtex4-1}

\draft % marks overfull lines with a black rule on the right

\usepackage{graphicx}
\usepackage{dcolumn}
\usepackage{bm}

\usepackage{color}

\begin{document}


\title{Hybridized electronic states in potassium-doped picene probed by \\soft x-ray spectroscopies}

\author{Hiroyuki Yamane}
\email[Electronic mail: ]{yamane@ims.ac.jp}
\affiliation{Institute for Molecular Science, Okazaki 444-8585, Japan}
\affiliation{The Graduate University for Advanced Studies, Okazaki 444-8585, Japan}

\author{Nobuhiro Kosugi}
\affiliation{Institute for Molecular Science, Okazaki 444-8585, Japan}
\affiliation{The Graduate University for Advanced Studies, Okazaki 444-8585, Japan}

\date{\today}

\begin{abstract}
The electronic structure of the unoccupied and occupied states of potassium (K)-doped and undoped picene crystalline films has been investigated by using the element-selective and bulk-sensitive photon-detection methods of X-ray absorption and emission spectroscopies. We observed the formation of the doping-induced unoccupied and occupied electronic states in K-doped picene. By applying the inner-shell resonant-excitation experiments, we observed the evidence for the orbital hybridization between K and picene near the Fermi energy. Furthermore, the resonant X-ray emission experiment suggests the presence of the Raman-active vibronic interaction in K-doped picene. These experimental evidences play a crucial role in the superconductivity of K-doped picene.
\end{abstract}



\maketitle

\section{Introduction}

Picene (C$_{22}$H$_{14}$) is a planar aromatic hydrocarbon molecule, which is regarded as small graphene's unit cell. Recently, a new organic superconductor was discovered for a potassium (K)-doped picene, which shows the transition temperature of $\it{T}\rm{_c}$ = 7$-$18 K, depending on the K concentration.\cite{Tc_1} This finding paves the way for the systematic examination of $\it{T}\rm{_c}$ due to the degree of freedom of molecules.\cite{Tc_2,Tc_3} In order to elucidate the nature of the superconductivity, detailed information on the electronic structure near the Fermi energy ($\it{E}\rm{_F}$) has been required. From theoretical investigations, it has been suggested for K-doped picene that the dopant (K) outer orbitals are weakly hybridized with the electronic states of picene close to $\it{E}\rm{_F}$, which plays a crucial role in the superconductivity.\cite{DFT_1,DFT_2,DFT_3} Besides the theoretical works, experiments on the electronic structure of K-doped and undoped picene have also been performed by using electron spectroscopies.\cite{PES_1,PES_2,PES_3,EELS_1,EELS_2} These experimental works show the formation of new electronic states near $\it{E}\rm{_F}$ upon K doping. However, detailed characteristics of doping-induced electronic states in K-doped picene have not yet been clarified due to the lack of element/site selectivity in the earlier experiments.

The X-ray emission spectroscopy (XES), a photon-in (excitation) photon-out (de-excitation) technique (Fig.~\ref{fig1}), gives the element/site-selective characterization of valence electronic states through highly localized core excitation and the excitation dynamics of materials with inherent bulk sensitivity by X-rays. Because of the rapid progress in detection efficiency and energy resolution of the spectrometer, the XES has been applied to the characterization of various materials.\cite{XES_1,XES_2,XES_3} Furthermore, the XES has been used for the detection of very small partial density-of-states and resonant inelastic scattering in various materials including organic molecules.\cite{XES_4}

In the present work, in order to understand the element-resolved bulk electronic structure of K-doped picene, the fluorescence-yield X-ray absorption spectroscopy (FY-XAS) and XES experiments were performed for crystalline films of undoped and K-doped picene (K$\it{_x}$picene, $\it{x}$ = 0, 3), wherein $\it{x}$ = 3 gives the highest $\it{T}\rm{_c}$ of 18 K.\cite{Tc_1} From the inner-shell resonant experiments, we observed the impact of the K-doping on the occupied and unoccupied electronic structure of picene, the orbital hybridization of picene with K, and the presence of the Raman-active vibronic interaction in K-doped picene.

\section{Experimental}

The FY-XAS and XES experiments were performed at a soft X-ray in-vacuum undulator beamline BL3U of the UVSOR facility, the Institute for Molecular Science.\cite{BL3U_1,BL3U_2,BL3U_3} The FY-XAS spectra were measured by using a micro-channel plate (MCP) assembly with a center hole.\cite{BL3U_4} In order to obtain the FY-XAS spectra, the retarding bias of $-$1 kV was applied to the mesh deflector, which is placed at just before the MCP detector [see, inset of Fig.~\ref{fig2}(b)]. The retarding bias of $-$1 kV shuts out the photoelectrons excited by not only the fundamental X-rays but also the false second-order X-rays generated from gratings. The XES spectra were measured by using a transmission-grating spectrometer, which adopts a free-standing transmission grating (5555 lines/mm), a Wolter type I mirror, and a normal-incident back-illuminated charge-coupled-device detector.\cite{BL3U_1,BL3U_2,BL3U_3} In the present work, the energy resolving power ($\it{h\nu}$/$\it{\Delta E}$) is set to 3000 for FY-XAS and 1000 for XES. All measurements were performed at 300 K. In order to avoid the radiation damage due to the long-time exposure of soft X-rays with 10$^{11}$ photons/sec, the X-ray irradiated spot was scanned at the rate of 20 $\it{\mu}$m/min during the XES measurements (60 min for 1 spectrum, in the present work).

For the qualitative reference of the experimental data, the molecular orbital calculations were performed by using the \verb|Gaussian 03| program package.\cite{G03}

The K$\it{_x}$picene ($\it{x}$ = 0, 3) samples were 100-nm-thick films, as prepared under UHV by the co-evaporation of picene (99.9\% purity, NARD institute Ltd., Japan) and K (SAES Getters S.p.A., Italy) onto the naturally oxidized $\it{n}$-type Si(111) substrate [SiO$_2$/Si(111)] at the deposition rate of 0.5 nm/min with the post-annealing at 370 K for 2 hours. The K concentration and the crystallinity of the K$\it{_x}$picene ($\it{x}$ = 0, 3) films were confirmed by FY-XAS and the CuK$\rm{\alpha}$ X-ray diffraction (XRD, Rigaku RINT-Ultima III).

\section{Results and Discussion}

Figure~\ref{fig2}(a) shows the specular CuK$\rm{\alpha}$ XRD spectra ($\it{\lambda}$ = 0.154 nm) of the undoped and K-doped picene films, wherein the abscissa is given by 2$\it{\theta}$ and the momentum transfer ($4\pi\lambda^{-1}\sin\theta$). The observed XRD peak positions for the present film sample agree well with those for the powder sample in the earlier study,\cite{Tc_1,XRD} indicating the validity of our sample crystallinity, that is, the K atoms are intercalated in the stacked picene molecules. The XRD spectrum for the undoped picene shows the (001), (002), and (003) peaks due to the higher order diffraction of the (001) plane in the Bragg's law of $2d\sin\theta = n\lambda$ with $n = 1, 2, 3$. On the other hand, the XRD spectrum for the K-doped picene is relatively complicated due to the presence of three K atoms for one picene molecule and their higher order diffraction. However, the dominant (001) peak is stronger more than 10 times than the other diffraction peaks. From the observed intensity distribution in the present XRD data, wherein the (001) peak around the momentum transfer of 4.65 nm$^{-1}$ is dominant, we found for both the undoped and K-doped picene films that the molecules on the SiO$_2$/Si(111) substrate form a (001) standing orientation dominantly, as illustrated in the inset of Fig.~\ref{fig2}(a), which is also supported by FY-XAS as discussed below.

Figure~\ref{fig2}(b) shows the polar angle ($\it{\alpha}$) dependence of the FY-XAS spectra of the undoped and K-doped picene films with the horizontally polarized X-rays. For the undoped picene film, the strong FY-XAS peaks appear at the incident photon energy ($\it{h\nu}\rm{_{in}}$) of 284$-$285 eV. According to the previous XAS studies on polycyclic aromatic hydrocarbons,\cite{XAS_PAH1,XAS_PAH2} the strong FY-XAS peaks at $\it{h\nu}\rm{_{in}}$ = 284$-$285 eV can be ascribed to the C 1s $\to$ lowest unoccupied molecular orbital (LUMO, $\pi^\ast$) and LUMO+1 ($\pi^\ast$) excitations. These $\pi^\ast$ excitation peaks are strongest at the normal incidence geometry ($\it{\alpha}$ = 0$^{\circ}$) and getting weaker at the grazing incidence geometry ($\it{\alpha}$ = 60$^{\circ}$). The change in the FY-XAS intensity with the incident angle can be explained by the transition probability for the C 1s $\to$ $\pi^\ast$ excitation with polarized X-rays.\cite{XAS_book} The present experimental evidence in FY-XAS indicates the standing molecular orientation. Upon K doping, the K L$_{2,3}$-edge peaks appear at $\it{h\nu}\rm{_{in}}$ = 295$-$301 eV, and the $\pi^\ast$ excitation peaks are weakened and broadened. Furthermore, the $\pi^\ast$ excitation onset is shifted to the lower $\it{h\nu}\rm{_{in}}$ side upon K doping. The K L$_{2,3}$ peak intensity, where the intensity ratio L$_3$:L$_2$ is not 2:1 but $\approx$1:1, indicates the presence of the orbital hybridization or covalent bonding. This is also supported by the presence of the strong satellite structure in the K L$_{2,3}$ peaks, which is ascribed to the charge transfer.\cite{XAS_K} Thus, the spectral change in the $\pi^\ast$ excitation peaks upon K doping can be explained by the electron transfer from K to picene. Note that, also for the K$_3$picene film, the $\pi^\ast$ excitation peak is getting weaker at the grazing incidence geometry, indicating the standing molecular orientation. By combining the XRD and FY-XAS results, we conclude that both the undoped and K-doped picene films on the SiO$_2$/Si(111) substrate form the (001) standing molecular orientation.

On the basis of the information on the crystalline structure and the unoccupied electronic states of the K$\it{_x}$picene ($\it{x}$ = 0, 3) films, we now discuss the occupied electronic states using XES. Here, according to the selection rule for the X-ray emission derived from the p$\rm{_z}$ $\to$ s orbital decay channel, the X-ray emission probability is higher along the perpendicular direction with respect to the p$\rm{_z}$ orbital axis. Therefore, in the case of standing $\pi$-conjugated molecules, the XES intensity is stronger near the surface normal direction as shown in the inset of Fig.~\ref{fig3}(b).

\begin{figure}
\includegraphics{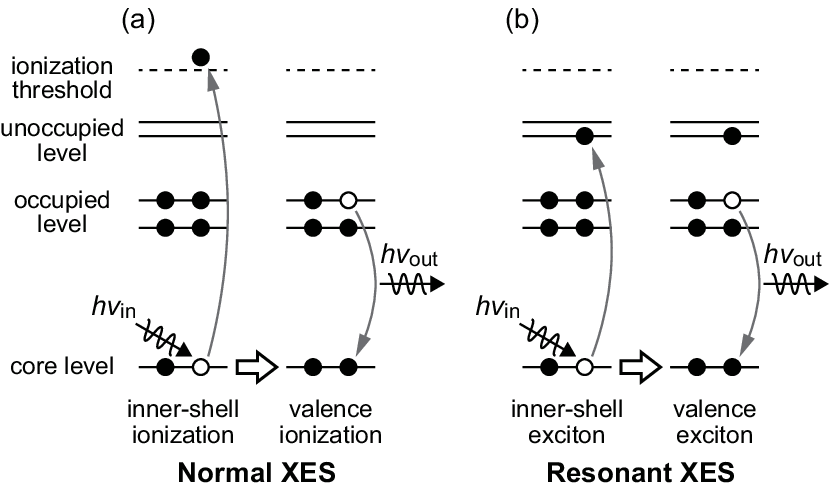}%
\caption{\label{fig1} Schematic representation of the normal (off-resonant) and resonant X-ray emission processes. (a) In the normal emission process, the core electron is excited above the ionization threshold energy, and then the valence electron decays into the core hole with X-ray emission. (b) In the resonant emission process, the core electron is resonantly excited into the unoccupied states, and then the valence electron decays into the core hole with X-ray emission.}
\end{figure}

\begin{figure}
\includegraphics{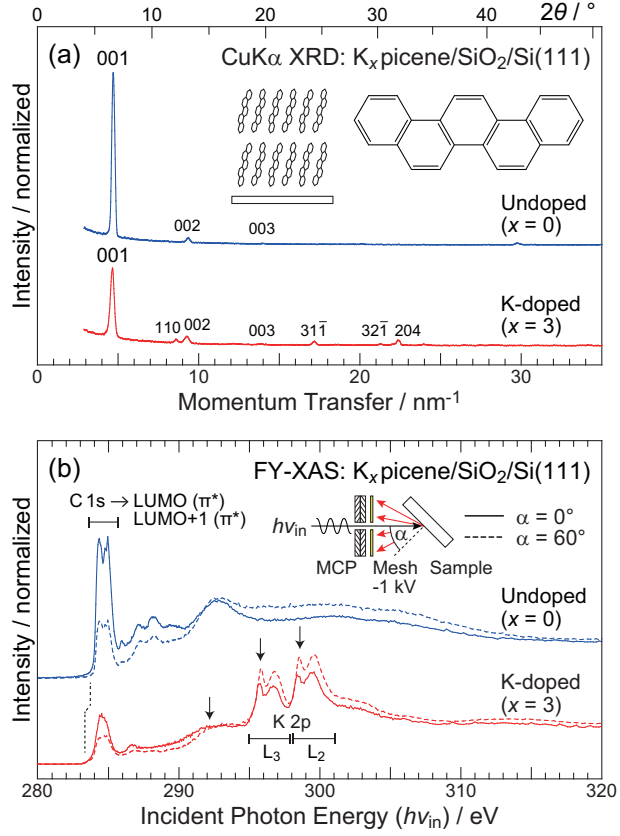}%
\caption{\label{fig2} (a) CuK$\rm{\alpha}$ XRD spectra ($\it{\lambda}$ = 0.154 nm, specular scan) for the undoped and K-doped picene films prepared on SiO$_2$/Si(111). The abscissa is given by 2$\it{\theta}$ and the momentum transfer ($4\pi\lambda^{-1}\sin\theta$) for convenience. Schematics of the picene molecule and its crystalline structure are shown in the inset. (b) FY-XAS spectra for the undoped and K-doped picene films prepared on SiO$_2$/Si(111), measured at the normal incidence ($\it{\alpha}$ = 0$^{\circ}$, solid line) and grazing incidence ($\it{\alpha}$ = 60$^{\circ}$, dashed line) geometries. The experimental geometry for the FY-XAS spectra is shown in the inset. Downward arrows indicate the $\it{h\nu}\rm{_{in}}$ position for the resonant XES spectra shown in Fig.~\ref{fig3}(b).}
\end{figure}

\begin{figure}
\includegraphics{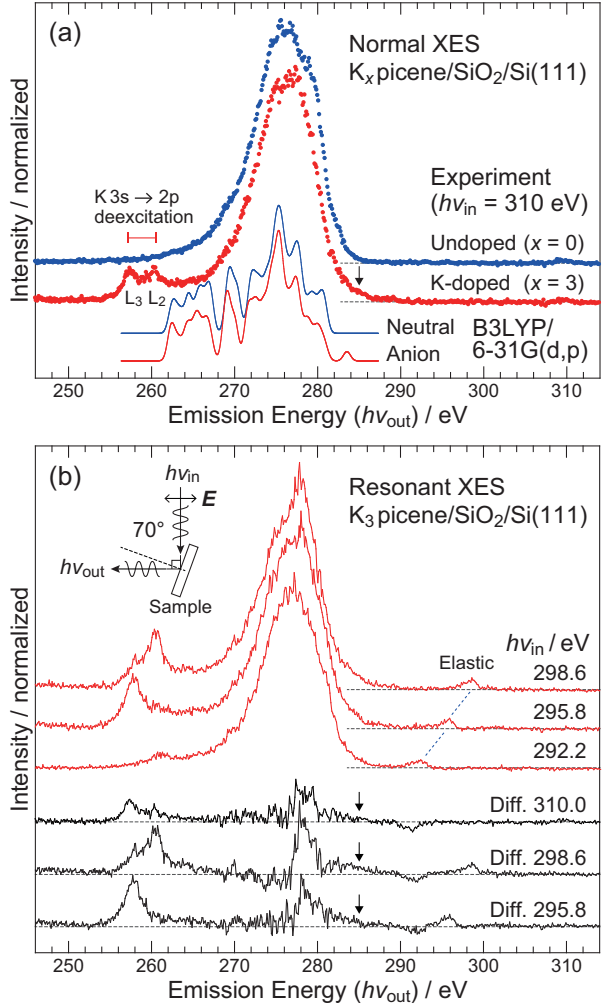}%
\caption{\label{fig3} (a) Normal (off-resonant) XES spectra for the undoped and K-doped picene films, measured at $\it{h\nu}\rm{_{in}}$ = 310 eV, and the calculated occupied electronic states for the neutral and anion picene molecule based on the density-functional theory using B3LYP with 6-31G(d,p) basis set. The calculated results are convoluted with the Gaussian broadening of 1.0 eV. (b) Resonant XES spectra for the K$_3$picene film near the K L$_{2,3}$-edge [cf. Fig.~\ref{fig2}(b)]. The difference XES spectra subtracted by the non-resonant spectra measured at $\it{h\nu}\rm{_{in}}$ = 292.2 eV are shown in the bottom. Inset shows the experimental geometry for the XES measurements.}
\end{figure}

Figure~\ref{fig3}(a) shows the normal (off-resonant) XES spectra, measured at $\it{h\nu}\rm{_{in}}$ = 310 eV, for the K$\it{_x}$picene ($\it{x}$ = 0, 3) films, wherein the abscissa is the emitted photon energy ($\it{h\nu}\rm{_{out}}$). The convoluted curves of the calculated occupied states for a picene molecule in the neutral and anionic states, performed by the density-functional theory using B3LYP\cite{B3LYP} with 6-31G(d,p) basis set with the Gaussian broadening of 1.0 eV, are shown for qualitative reference. In the normal XES, the final state is the one hole (1h$^+$) state following the dipole transition from the inner-shell ionized states to the valence ionized states as shown in Fig.~\ref{fig1}(a). Therefore, the normal XES spectra reflect partial density-of-states of materials. For both the undoped and K-doped picene films, the broad emission features appear at $\it{h\nu}\rm{_{out}}$ = 270$-$280 eV prominently. This prominent emission feature shows the moderate change upon K-doping. Such a doping effect more clearly appears as new peaks at $\it{h\nu}\rm{_{out}}$ = 257 eV (L$_3$) and 260 eV (L$_2$), which are ascribed to the K 3s $\to$ 2p deexcitation. At the higher $\it{h\nu}\rm{_{out}}$ around 285 eV, which is close to $\it{E}\rm{_F}$, the XES intensity slightly increases upon K doping as indicated by the downward arrow in Fig.~\ref{fig3}(a). On the other hand, in the calculated electronic structure, the density-of-states at $\it{h\nu}\rm{_{out}}$ = 280 eV in the anionic state is weaker than that in the neutral state. Furthermore, a new peak, which originates from the filling of the former LUMO, appears near $\it{E}\rm{_F}$ in the anionic state. These evidences are comparable to the experimental results, and therefore, the observed XES feature at $\it{h\nu}\rm{_{out}}$ = 285 eV for the K-doped picene film might be ascribed to the former-LUMO-derived level due to the electron transfer from K to picene.

In order to analyze the doping-induced XES feature near $\it{E}\rm{_F}$ in more detail, we applied the resonant-excitation measurements. In the resonant XES, the final state is the one hole and one electron (1h$^+$1e$^-$) state as shown in Fig.~\ref{fig1}(b), and it is described as a resonant X-ray Raman process. Furthermore, the resonant inner-shell excitation particularly enhances the cross section for the inelastic X-ray scattering, which involves the element-specific intrinsic excitations of materials. Figure~\ref{fig3}(b) shows the selected resonant XES spectra of the K$_3$picene film, measured at various $\it{h\nu}\rm{_{in}}$. Before the K L$_3$ resonance ($\it{h\nu}\rm{_{in}}$ = 292.2 eV), the XES feature at $\it{h\nu}\rm{_{out}}$ = 285 eV (around $\it{E}\rm{_F}$) weakly appears. This feature is enhanced at the K L$_3$ and L$_2$ resonances ($\it{h\nu}\rm{_{in}}$ = 295.8 and 298.6 eV, respectively), as well as the K 3s $\to$ 2p peak. These spectral change by the resonant excitation can be found more clearly in the difference spectra subtracted by the non-resonant spectra measured at $\it{h\nu}\rm{_{in}}$ = 292.2 eV. In the difference spectra shown in the bottom of Fig.~\ref{fig3}(b), the intensity of the XES feature around $\it{E}\rm{_F}$, indicated by the downward arrow, is almost the same as that of the elastic peak when $\it{h\nu}\rm{_{in}}$ = 295.8 eV and 298.6 eV, which are the resonant excitation condition. This XES feature around $\it{E}\rm{_F}$ is getting weaker when $\it{h\nu}\rm{_{in}}$ = 310.0 eV due to the off-resonance effect. This experimental evidence indicates the presence of the K-derived orbitals around $\it{E}\rm{_F}$, which definitely supports the theoretical prediction for K$_3$picene that the K 4s electrons are donated to unoccupied states derived from the LUMO and LUMO+1 of picene with orbital hybridization. Based on the present FY-XAS and XES observations, in order to achieve higher density-of-states around $\it{E}\rm{_F}$ upon K doping, the degenerate unoccupied electronic states in the crystalline phase might be important.

On the other hand, the elastic peak in the resonant XES spectra shows the linear dispersion with $\it{h\nu}\rm{_{in}}$. Although the elastic peak weakly appears, it shows an asymmetric lineshape with a low-energy tail structure. Such a tail structure in the resonant XES spectra has been observed also for other molecular solids of mixed Watson-Crick base pairs (Ref.~\onlinecite{RIXS_1}) and organometallic compounds (Ref.~\onlinecite{RIXS_2}). In the case of mixed Watson-Crick base pairs, presence of the ultrafast charge motion accompanied by the vibronic interaction has been suggested by resonant photoemission experiments.\cite{RPES_1} Furthermore, it has been examined that such an asymmetric elastic peak with the low-energy tail structure is observable depending on the crystalline structure, wherein the ultrafast motion of the excited electron exists in the de-excitation process.\cite{RIXS_2} In the case of amorphous films, the ultrafast charge motion does not occur, and thereby the elastic peak shows a Gaussian-type symmetric lineshape.\cite{RIXS_2} Judging from these studies, the observed tail structure might be dominated by intrinsic phenomena. As the possible origin of the observed tail structure, the recombination emission\cite{RIXS_3} due to the presence of the Raman-active vibronic interaction in K$_3$picene can be considered.

\section{Conclusion}

In conclusion, we have studied the unoccupied and occupied bulk electronic states of K$\it{_x}$picene ($\it{x}$ = 0, 3) by using the element/site-selective X-ray spectroscopies. By combining the FY-XAS and XES results, we could observe the formation of the doping-induced electronic states in K$_3$picene, which shows the evidence of the orbital hybridization between picene (C 2p) and potassium (K 4s) near $\it{E}\rm{_F}$. This observation is the direct experimental proof for the importance of the degenerate unoccupied states in the crystalline phase to characterize the superconductivity of K-doped aromatic hydrocarbon molecules. Furthermore, the elastic peak in the resonant XES spectra of K$_3$picene shows the asymmetric lineshape with the low-energy tail structure, which suggests the presence of the Raman-active vibronic interaction in K-doped picene.

\begin{acknowledgments}
The authors are grateful for the kind supports by the staff of the UVSOR facility. The present work was partly supported by Grant-in-Aid for Challenging Exploratory Research (No. 24656022) and Grant-in-Aid for Scientific Research (A) (No. 23245007).
\end{acknowledgments}

\clearpage

\clearpage

\clearpage

\end{document}